\documentclass[preprint,superscriptaddress,pre,a4]{revtex4} 
\usepackage{latexsym}
\usepackage{times}
\usepackage{graphicx}
\usepackage[tight,FIGTOPCAP]{subfigure}

\usepackage{color}

\newcommand\TOPFIX{\rule{0pt}{2.7ex}}
\newcommand\BOTTOMFIX{\rule[-1.2ex]{0pt}{0pt}}

\usepackage[T1]{fontenc}
\usepackage{amsmath}
\usepackage{amssymb}
\usepackage{amsfonts}
\usepackage{amsthm}
\usepackage{multirow}

\color{black}

\begin{document} 

%
\author{A.~L.~Sellerio}
\affiliation{D\'epartement de Physique, Ecole Polytechnique
  F\'ed\'erale de Lausanne, CH-1015 Lausanne, Switzerland}
\author{B.~Bassetti}
\affiliation{Universit\`a degli Studi di Milano,
  Dip.  Fisica. Via Celoria 16, 20133 Milano, Italy}
\altaffiliation{and I.N.F.N. Milano, Italy.}
\author{H.~Isambert}
\email[ e-mail address: ]{herve.isambert@curie.fr}
\affiliation{UMR 168 / Institut Curie, 26 rue d'Ulm 75005 Paris, France}
\author{M.~Cosentino Lagomarsino} 
\email[ e-mail address: ]{Marco.Cosentino-Lagomarsino@unimi.it}
\affiliation{Universit\`a degli
  Studi di Milano, Dip.  Fisica.  Via Celoria 16, 20133 Milano, Italy}
\altaffiliation{and I.N.F.N. Milano, Italy.}



\title{A comparative evolutionary study of transcription networks.}


\keywords{ transcription networks / topology / feedback / hierarchy /
  gene duplication / evolution  }

\begin{abstract}
  We present a comparative analysis of large-scale topological and
  evolutionary properties of transcription networks in three species,
  the two distant bacteria \emph{E.~coli} and \emph{B.~subtilis}, and
  the yeast \emph{S.~cerevisiae}.  The study focuses on the global
  aspects of feedback and hierarchy in transcriptional regulatory
  pathways.  While confirming that gene duplication has a significant
  impact on the shaping of all the analyzed transcription networks,
  our results point to distinct trends between the bacteria, where
  time constraints in the transcription of downstream genes might be
  important in shaping the hierarchical structure of the network, and
  yeast, which seems able to sustain a higher wiring complexity, that
  includes the more feedback, intricate hierarchy, and the
  combinatorial use of heterodimers made of duplicate transcription
  factors.
\end{abstract}


\maketitle

\section{Introduction}

Cells need constant sensing of environmental changes and internal
fluxes, and correct response to these external and internal stimuli
through the simultaneous expression of a large set of genes.  The
basal mechanism that performs this task is transcriptional
regulation. Depending on species, context and specific function, this
can involve simple interactions or complex signaling cascades, but in
general it involves a large number of genes~\cite{RS93, PWG99, PC00,
  BB04, UKZ05, BBA07}.
For this reason, it is necessary to characterize this regulatory
process from a global, or "network" point of view. To this aim,
transcriptional regulation networks are defined starting from the
basic functional elements of transcription~\cite{BLA+04}.  This
information is represented as a directed graph, usually identifying
each gene transcript and their protein products with an unique node,
and each regulatory interaction with a directed edge \( A \rightarrow
B \) between the node \(B\) (the target gene) and the node \(A\) (the
gene coding for a transcription factor (TF) that has at least one
binding site in the cis-regulatory region of \(B\)). A transcription
factor regulating its own expression is called an autoregulator (AR).
With this definition, the interaction graph structure is given by
large-scale and collections of small-scale experiments~\cite{SMM+02,
  SSG+06, LRR+02, HGL+04}.

A basic way to understand the architecture of transcription networks
it to consider their topology. Topological analysis is able to capture
functional properties, and important architectural features of the
network~\cite{MIK+04, WtW04, TB04, MBZ04, MKD+04, YG06, CLJ+07}.
Examples of topological properties are the so-called network
\emph{motifs}~\cite{SMM+02}, or the degree distribution of the
connectivity of the nodes~\cite{IMK+03}.
%
%
These observations need to be compared with suitable null models.  For
example, ``network motifs'' are subgraphs for which the probability
\(P\) of appearing in a suitably randomized network an equal or
greater number of times than in the empirical network is lower than a
given cutoff value.  
For a meaningful comparison, randomized networks are taken with the
same single-node characteristics of the empirical network.  Usually,
each node in the randomized networks is constrained to have the same
number of incoming and outgoing edges (degree distribution) as the
corresponding node has in the empirical
network~\cite{MSI+02,FBJ+07}. Considering more large-scale properties,
the known transcription networks possess a hierarchical feedforward
layered structure~\cite{MBZ04,YG06}, and often feedback is mainly
limited to a rather large set of autoregulations~\cite{THP+98}.
The perspective on the topology transcription networks is enriched by
taking an evolutionary point of view.  Evolution of a transcription
network is driven by three main biological mechanisms: (\textbf{i})
gene duplication, (\textbf{ii}) rewiring of edges by
mutation/selection of TF/DNA interactions and (\textbf{iii})
horizontal gene transfer. The first mechanism has been shown to play a
substantial role, although the extent to which it can shape the
network is debated~\cite{BLA+04, TB04, CW03, DMA05, MBV05}.  For
example, it has been shown that network motifs do not emerge from
duplication events~\cite{MBV05, BLA+04}, while other topological
properties have arisen from gene duplications~\cite{BLA+04, BBI+06}.
We have previously considered from this viewpoint the properties of
hierarchy and feedback in the \emph{E.~coli} network~\cite{CLJ+07},
finding that gene duplication can be held responsible for the
preservation of self-regulations, and for the ``shallow'' layered
organization, which one can hypothesize to optimize the time
constraints for the production of targets.

In this paper, we present a comparative study of transcription
networks, which extends the analysis on \emph{E.~coli}, and considers
also the evolutionarily distant bacteria \emph{B.~subtilis}, and two
different data sets for the eukaryote \emph{S.~cerevisiae}.
From comparison of these data we distinguish between unifying and
distinct features of the three networks.
In particular, while in the two bacteria feedback loops involve few
nodes and the number of hierarchical layers is minimal, yeast shows
more feedback and a more complex hierarchy of transcription factors.
Autoregulatory interactions are always abundant, although their role
appears to be different between yeast and the bacteria.  The fraction
of self-regulators is above \(50\%\) in the bacterial data sets,
dropping around \(10\%\) for \emph{S.~cerevisiae}.
If we take into account the effects of evolution on topology, a
richer, more complex scenario appears.  Our analysis is focused on the
mechanism of gene duplication, and is based on a network growth model
which considers this drive to be the only one present.  With this
method it is also possible to infer on other mechanisms indirectly.
Our main findings are the following.  (\textbf{i}) Duplications play
an important role in evolution of the TN: the relative abundance of
simple network subgraphs stemming from duplication is evident in all
data sets.  (\textbf{ii}) Gene duplications shape the degree sequences
and the hierarchy of the network, as predicted by a
duplication-divergence model. (\textbf{iii}) The feedback core in the
yeast network may be shaped by duplications of existing feedbacks
(\textbf{iv}) The yeast network tends to form heterodimeric TF pairs
from duplicates, which seem to be forbidden in \emph{E.~coli}, possibly
because of the same selective pressure erasing crosstalks from
duplicate ARs.

\section{Feedback and hierarchy. Topological Evaluation.}

Feedback is present in our network if closed directed paths exist.  We
quantify the amount of feedback with the ``leaf-removal'' algorithm.
This decimation algorithm iteratively removes the input and output
tree-like components (\emph{i.e.} parts without loops) of a directed
graph~\cite{CLB06}.  The outcome can be either the empty graph or a
non-empty subgraph of the original.  In the former case, the whole
network is tree-like. In the latter, we say that the leftover subgraph
represents the feedback \emph{core} of the network.
The size of this core (number of nodes or edges, with respect to the
complete network) can be used as a measure of the amount of feedback.

The reverse, or complementary, of feedback is \emph{hierarchy}.
Neglecting self-regulations, we define \emph{roots} the nodes which
are not regulated by other nodes, and \emph{leaves} the nodes which do
not regulate other nodes.  Starting from the roots of the graph we can
define hierarchical layers considering the relative ordering of nodes
in a chain of regulatory interactions.  (\textbf{i}) If we consider a
tree-like graph (or a tree-like subset of it), each iteration of the
leaf-removal algorithm removes a number of nodes, either roots or
leaves, and the edges pointing from or to these nodes. By definition,
the removed nodes do not interact with each other, and can be thought
to form a distinct computational ``layer''.  In this case, the
leaf-removal algorithm described above naturally defines a hierarchy
between the nodes, through the longest path to a root or to the
feedback core upstream of a given node.  In the general case, we
define the number of layers in the hierarchy by the number of
iterations of the leaf-removal algorithm needed to remove all the
nodes outside of the feedback core (thus as a sum of layers downstream
and upstream of the feedback).  (\textbf{ii}) Hierarchical layers can
be defined in a similar way through the longest open chain of
regulators that each of their members share.  In a longest-path
hierarchy, and in absence of feedback, members of layer one are
regulated by at most themselves.  Members of layer two are regulated
by a chain of one, and no more, nodes and possibly themselves, and so
on.  This definition, though conceptually similar to that given by the
leaf removal, is computationally demanding since the algorithm
represents a NP-complete problem. We did not use this definition in
our analysis.  (\textbf{iii}) An alternative definition
(computationally easier) is given by the shortest open directed paths
between nodes.  The number of layers is computed considering the
longest among the shortest paths from any pairs of nodes (which can be
found in polynomial time, for example with the Dijkstra
algorithm~\cite{Dij59}). Shortest paths measure the minimal number of
intermediary transcriptional interactions required for a signal from a
transcription factor to reach a given target downstream, which can be
interpreted as hierarchical layers. On the other hand, a
straightforward univocal definition of hierarchical layers using
shortest paths is difficult to produce~\cite{YG06}.

\begin{figure}[ht]
  \centering
  \includegraphics[width=0.7\textwidth]{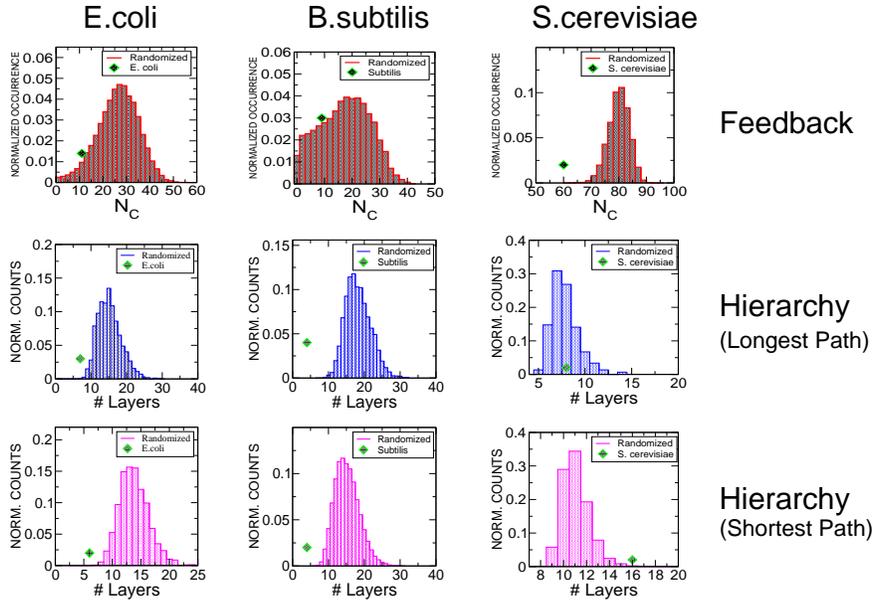}
  \caption{Feedback and hierarchical layers for the three networks of
    \emph{E.coli} (RegulonDB 5.5) \emph{B.~Subtilis} (DBTBS) and
    \emph{S.~Cerevisiae} (Balaji data set), compared to randomized
    instances.  The data on \emph{E.~coli} are compatible with older
    data sets. The two data sets on yeast do not give compatible
    results; data refer to most recent one. Top: Number of
    nodes \(N_{C}\) in the feedback core.  Middle: Number of
    leaf-removal layers for the empirical networks and randomizations
    having the same or lower \(N_{C}\) as the empirical ones (to
    improve the size of the sample, we show \(N_{C} <70\) for the case
    of yeast, however, the results do not change for lower
    thresholds). Releasing the constraint on \(N_{C}\) gives a weaker
    signal for the bacteria.  Bottom: Number of layers in a
    shortest-path hierarchy, without constraints on \(N_{C}\) for the
    randomized cases. A constraint for the empirical \(N_{C}\) gives a
    stronger signal.}
  \label{fig:topology}
\end{figure}

In order to quantify the significance of feedback and hierarchical
properties of the three transcription networks (five data-sets), we
compared them with random ensembles of networks having the same degree
sequences, i.e. conserving the number of incoming and outgoing edges
for each node.
For all the bacterial data sets, we find consistently the following.
The feedback core is slightly smaller than the typical random case,
and the number of layers is minimal. This is true both for a longest
path and a shortest path hierarchy. Data are shown in the first two
columns of Fig.~\ref{fig:topology}.  On the other hand, the case of
\emph{S.~cerevisiae} is more complex. First, we observe that the two
data sets we analyzed are not consistent with each other. We believe
this to be due to their large differences in size: the older Guelzim
data set~\cite{GBB+02} contains approximately \(800\) interactions,
while the more recent Balaji data set\cite{BBI+06} contains around
\(13000\).  Considering the Guelzim data set, the almost tree-like
topology does not differ from randomizations in any of the observables
considered above.  For the Balaji data set, we find contrasting trends
compared to the bacterial networks.  The rather large feedback core
involving \(60\) nodes and \(207\) interactions is far larger than the
one found in the two bacterial networks.  On the other hand, it is
significantly smaller than the typical feedback core found in
randomizations.  Interestingly, the number of leaf-removal layers
falls in the average, in contrast with the trend observed in bacterial
networks. Even more surprisingly, the number of shortest-path layers
in the yeast network greatly exceeds randomizations, in strong
contrast with the bacterial data-sets. Data are shown in the third
column of Fig.~\ref{fig:topology}.  More in detail (Supplementary
Fig.~\ref{fig:yeast_shpdist}, the lengths of these shortest-paths have
a Poisson-like distribution in both the empirical graphs and their
randomizations, but for the yeast network the tails of their
distributions are significantly shifted between the two.

A possible rationale for these observations is the importance, in
bacteria, of minimization of the time-scales for the production of
target genes, for example with structural connotation. Since each
transcriptional step of an expression program takes an estimated time
of the order of one cell cycle~\cite{REA02}. The same argument might
also explain the abundance of self-repressed transcription factors,
which reach steady-state expression in shorter
times~\cite{REA02}. This kind of pressure is possibly released from
yeast due to a more efficient organization of gene expression and
degradation.

\section{Evolutionary Duplication-divergence Growth}

\paragraph*{Model.}

We use a simple duplication-divergence model for the growth of the
transcription network~\cite{tesikirill, EI-06, CLJ+07} to guide the
data analysis. The model is formulated as follows. At each step all,
or a fraction of the nodes are duplicated. The duplicate nodes inherit
all the in- and outgoing edges of the original (ancestral) node. In
other words one supposes that before divergence, the binding sites on
the proteins and the regulatory regions on DNA remain identical.
Subsequently, the new edges are removed with a certain probability,
that might depend on the status of the node (regulator or target,
subject or regulating a new or an old gene, etc.).

A qualitative study of the model leads to the following schematic
results\cite{tesikirill}.
First, by definition there is no possibility of \emph{de novo}
addition of edges by rewiring. Hence no edges can be created from an
homology class that initially does not regulate another one.  In the
simplest case, if feedback is initially absent, it cannot arise
spontaneously.  Moreover, no hierarchical layer can be added to the
network, and duplicates selected for fixation will lie in the same
layer.
On the contrary, in the presence of feedback, and ARs in particular,
the model behaves differently. Duplication of ARs can give rise to
higher order feedback and to new ARs. However, this feedback will be
strictly confined among members of the same homology class.
%
Analogously, the degree distribution is piloted by the degree
distribution of duplicates. The in- and out- degree distributions can
be decoupled by choosing different removal probabilities for old/new
and new/old edges. One can obtain power-law out-degree and compact
in-degree distributions that are in close relation with the
phylogenetic conservation of the network proteins~\cite{EI-06,EI07}:
conserved nodes {\em must} exhibit at least one between the in- and
out-degree sequence with a scale-free distribution, while proteins
with only exponential degrees {\em cannot} be phylogenetically
conserved under a general duplication-divergence model. According to
this model, bacterial transcription networks with scale-free
out-degrees and exponential in-degrees are thus consistent with a
phylogenetically conserved set of transcription factors and a
non-phylogenetically conserved set of target genes (possibly due to
abundant horizontal tranfers of target genes).

In a realistic situation, the above observations may not apply because
of the possibility of edge rewiring, that may be able to shuffle the
hierarchy, create new feedback, affect the degree distribution, and
mix edges among homology classes. Note, however, that some regulatory
edges between apparently different homology classes may have also
resulted from very old edges within ancestral homology classes that
are no longer classified as single homology classes due to their
gradual divergence. Besides, while actual edge rewiring {\em may} have
occurred after homology class separation, it is clear that edge
deletion {\em must} typically happen after gene duplication (or else
extant regulatory networks would be densely cross-edged graphs.)
Hence, this model, which focuses on the necessary deletion of
duplicated edges, can assess the specific role of gene duplication and
edge deletion in the growth of the network and also underline their
possible shortcomings to delineate the role of other evolutionary
processes such as edge rewiring.

\paragraph*{Data Evaluation.}

Having in mind the above qualitative results we examine the
topological roles in empirical networks of nodes coming from the same
common ancestor.
We define proteins that are likely to share a common ancestor through
structural domain architectures~\cite{TB04, MBT03, BD92}.
These domains allow for the definition of larger classes than sequence
comparison alone~\cite{TB04}.  The database enables to associate an ordered
sequence of domains, or ``domain architecture'' to each protein.
We define protein homologs as proteins whose domain architectures are
identical neglecting domain repeats. This corresponds to a
conservative view of homology where no new domains are acquired or
lost after duplication.  The results are tested using different
definitions of homologs~\cite{sellerio002}.  Homology allows us to
construct sets, or classes, of proteins which have then supposedly
evolved from a common ancestral gene.
We have analyzed the distribution of regulatory edges between and
within classes of the likely duplicate genes.  The statistical
significance of the analysis in terms of homology classes is
established~\cite{BLA+04} by comparison with random shuffling of genes
(TFs and TGs separately) between classes.  This analysis is limited by
the number of nodes for which homology classes can be constructed, and
thus less sensitive to the particular data set.

\begin{figure}[htbp]
  \centering
  \includegraphics[width=0.7\textwidth]{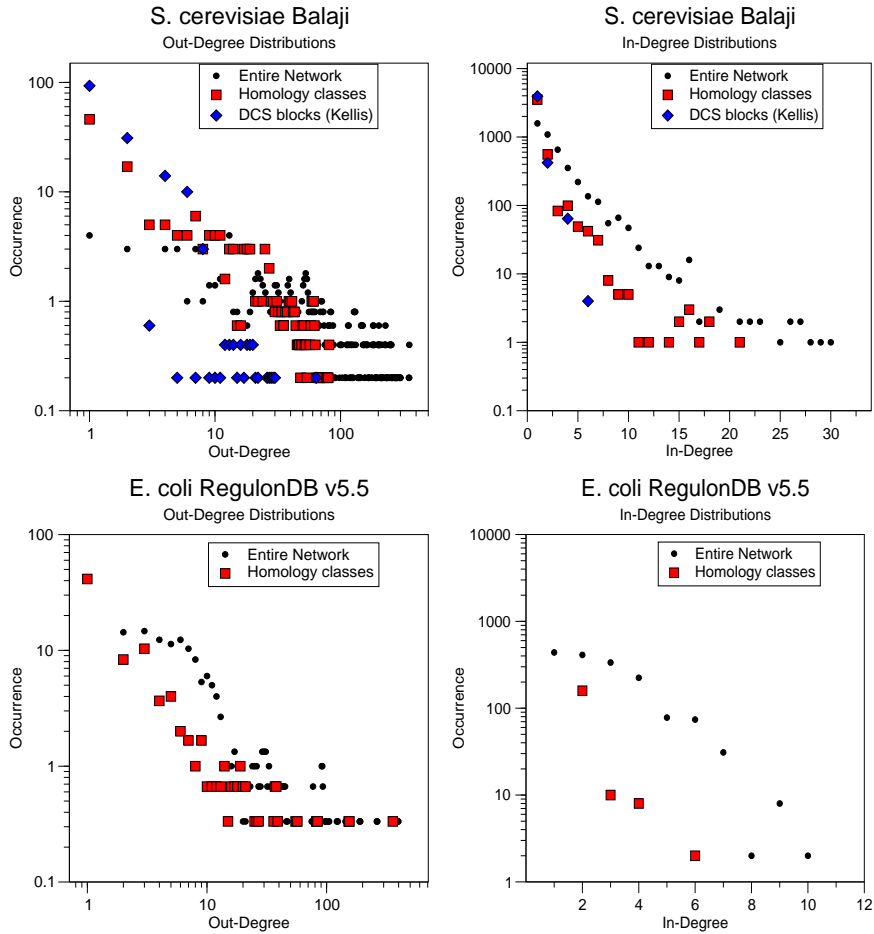}
  \caption{In- and out-degree distributions of the entire
    \emph{S.~cerevisiae} (Balaji data set) and \emph{E.~coli} (RDB
    5.5), compared to those obtained from homolog TGs and TFs.
    Homology is assessed with domain-architecture homology classes,
    and in the yeast, through doubly conserved genes in whole genome
    duplication.  The in-degree shows good agreement (right panel)
    with the prediction of the duplication model.  For the out-degree
    distribution, instead, data for the \emph{S.~cerevisiae} (top
    left) for the whole network seems to be wider than the homology
    classes, suggesting that processes other than duplication concur
    in shaping the out-connectivity of transcription factors. Some
    curves have been smoothed out by averaging nearby bins in order to
    enhance the visibility of the tails.}
  \label{fig:degree}
\end{figure}

In all cases we find that the motifs of duplicated TGs, regulated by a
single or duplicate TFs, and duplicate TFs regulating a common target
(Fig.~\ref{fig:degree}), are significantly overrepresented~\cite{TB04},
which can be seen as a validation of the model.  The signal for this
is smaller in the smaller data sets (such as \emph{B.~subtilis}),
because of poor statistics.

\paragraph*{Degree sequences and duplications. }
A first question to address is whether gene duplications are able to
pilot the observed degree sequences of the network, as predicted by
the model.
For the case of yeast, one can perform this test on the above defined
homology classes, and also, on the doubly conserved genes in the yeast
whole-genome duplication proved by Kellis \emph{et al.}~\cite{KBL04}.
Precisely, one can measure the degree distributions restricted to
targets or regulator falling in the same homology class, and verify
whether it scales in a similar way as the unrestricted distribution.

Our results, considering the Balaji data set for yeast and RegulonDB
5.5 for the \emph{E.~coli}, are shown in Fig.~\ref{fig:degree}.  The
homologs follow a qualitatively similar distribution to the entire
network, exponential for the in-degree, and power-law-like for the
out-degree.
For the in-degree distributions we find a better agreement between the
decay of the entire network and the restriction to homology classes.
This confirms the general idea that the degree distributions are
strongly dependent upon duplications.
However, in the case of yeast, the out-degree of the entire network
seems to follow a broader distribution than that of the duplicates
(top-left panel of Fig.~\ref{fig:degree}). This can be seen as an
indication that other processes, such as rewiring, concur in defining
the targets of a TF.
The behavior of the duplicates from the whole-genome duplication is
conditioned by the small size of the sample, and its degree
distribution drops with a faster decay for both the in- and the
out-degree.

\paragraph*{Duplication of Autoregulatory Circuits.}

Our analysis shows that the role of autoregulatory interactions is
rather different in the bacterial data sets compared to yeast.
The fraction of self-regulated TFs is above \(50\%\)
in all three bacterial data sets, while being around \(10\%\) only
in the two \emph{S.~cerevisiae} data sets.
If we measure directly the distribution of ARs in classes, in the
\emph{E.~coli} datasets we find a consistent signal that the
population of ARs in homology classes is more dense and more variable
than in randomized instances~\cite{CLJ+07}. This is a direct evidence
of duplication and conservation of ARs, and agrees with the
conservation of a hierarchical structure during evolutionary growth.
In \emph{B.~subtilis} and yeast, we find that the population of ARs in
classes falls closer to (although systematically above) the average of
the randomized samples. This means that, in these cases, the pure
analysis of AR population in homology classes is not sufficient to
assess that ARs in the same class are the result of duplication and
inheritance of self-edges.  In the smaller data-sets (such as
DBTBS~\cite{MNO+04}) this absence of signal could be likely due to
small size of the sample, so it is possible that this is another trend
differentiating bacteria from eukaryotes.

\begin{figure}[htbp]
  \centering
  \includegraphics[width=0.4\textwidth]{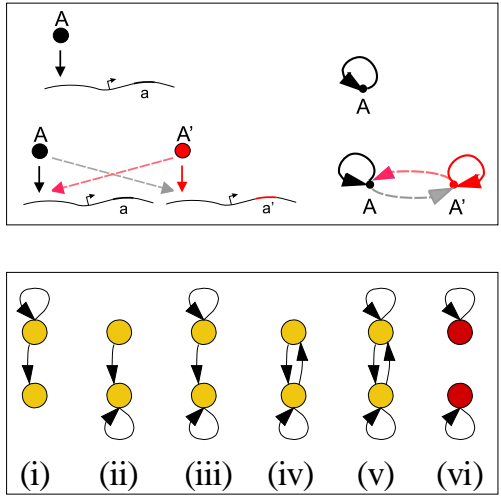}
  \caption{ Top: scheme of the possible circuits stemming from the
    duplication of a transcriptional self-regulation after partial
    inheritance of interactions.  Bottom: network subgraphs compatible
    with AR duplication and inheritance of crosstalks within homology
    classes. The circles represent homologous nodes.}
  \label{fig:arsub}
\end{figure}

\begin{table}
\centering
 \begin{tabular}{|c|c|c|c|}
 \hline
\(a\) \TOPFIX\BOTTOMFIX & Empirical & Randomization & Z-score\\ 
 \hline
\emph{E.~coli} (RDB 5.5)  & \(0.23\)  & \( 0.10  \pm  0.082 \)   & \(+1.6\) \\
\emph{B.~subtilis}    & \(0.03\)  & \( 0.0042 \pm 0.0073 \)  & \(+3.55\) \\
\emph{S.~cerevisiae} (Balaji) & \(0.13\) & \( 0.020 \pm 0.039 \) & \(+2.79\) \\
\hline
\end{tabular}
\caption{Subgraphs compatible with AR duplications, containing crosstalks
  (Fig.~\ref{fig:arsub}). The observable \(a\) measures the ratio
  between subgraphs and edges in the network, and is defined in the
  text.}
\label{table:motifs}
\end{table}

On the other hand, in all the data sets we find evidence for the
relevance of AR duplications if we consider the distribution of the
subgraphs that can stem from AR duplication (Fig.~\ref{fig:arsub})
within homology classes (Table~\ref{table:motifs}).  These subgraphs
are compatible with AR duplication conserving crosstalks.
In order to quantitatively evaluate the relevance of the subgraphs, we
define an observable that counts the occurrence of the following
events
\begin{enumerate}
\item An AR regulates a homologous, non-self-regulatory transcription
  factor.
\item An AR is regulated by a homologous, non-self-regulatory
  transcription factor.
 \item Two homolog ARs possess reciprocal cross regulation.
\end{enumerate}
We indicate with \(N_c\) the total number of homology classes,
with \(M_i\) the total number of the subgraphs of the
three kinds found in homology class \(i\), and by \(L_{i}\) the number
of edges in the class, and define
\[
a = \frac{1}{N_c} \sum_{i=1}^{N_c} \frac{M_i}{L_i} \quad .
\]
This observable represents the ratio of subgraphs to the total number
of (self- and non-self-) edges present in the homology class.

We find that, although the fixation of cross-talks within homology
classes is typically less frequent that the retention of self-edges in
bacteria~\cite{CLJ+07}, the AR duplication subgraphs are
systematically overrepresented, as shown by the positive signal. 
Specifically (see Supplementary Table~\ref{table:subgraphs_AR}), in
\emph{E. coli} and \emph{B. subtilis}, this signal is dominated by
duplicated self-interactions followed by loss of crosstalks, while for
the \emph{S. cerevisiae} network, subgraphs including crosstalks are
more abundant, compatibly with the observed more complex
hierarchy~\cite{CLJ+07}. In addition, in the significantly
overrepresented AR duplication subgraphs (type \emph{ii} in figure
\ref{fig:arsub}), only one of the two possible self-regulations of
duplicate TFs survives, indicating that ARs do not proliferate by
duplication.
This observation indicates that duplications have been important to
shape the non-self-regulatory edges within homology classes, thereby
redefining the TF hierarchy, besides leading to the fixation of new
ARs in the bacterial networks.
This is especially the case in yeast, indicating that circuits with
crosstalks or feedback stemming from AR duplication are not
negligible, and might have contributed to the build-up of the existing
feedback.

\paragraph*{Interaction Across Homology Classes and Evolution of the
  Feedback Core.}

To gain further insight into this point, we evaluated the distribution
of interactions within and across homology classes.  Considering the
transcription factors homology classes, we constructed a ``collapsed''
weighted network as follows. The nodes are homology classes, and
weighted oriented edges represent the number of members of a class
regulating members of another class.
In absence of rewiring, this would be a likely ``ancestral'' TF-TF
network, inherited by duplication of ``primitive'' self- and
heterologous edges. Our model dictates that this ancestral graph can
only be hierarchical or maintain the same amount of feedback of the
initial configuration.  In particular, if an ancestral edge is sent
out from a class to a second one and the reverse is not true, there
will be no edges coming from the second class to the first in the
evolved network. This corresponds to an {\em asymmetric} regulatory
control of the first homology class over the second homology class.
Conversely, the appearance of symmetric regulations may be a signature
of rewiring, if one assumes that retaining both ancestral crosstalks
produced by the duplication of an ancestral AR was then as unlikely as
it appears to be between recent AR duplicates.  Hence, symmetric
regulation between homology classes in the extant network actually
suggests that the likely initial asymmetry conserved by duplication
was progressively smoothed down by edge rewiring or other evolutionary
mechanisms with an equivalent effect.

We define the observable \(R\), which represents a measure of the
strength of asymmetric regulation across the classes.
We indicate as \(L_{ij}\) the number of regulatory edges between
homology class \(i\) and class \(j\). Let then \(C_{i}\) represent
the number of elements in homology class \(i\).
Then we define
\[
 R = \sum_{ {i \neq j }}
 \frac{\vert L_{ij}-L_{ji}\vert}{\sqrt{C_{i}^{2} + C_{j}^{2} }} \quad ,
\]
where the sum is performed over all the pairs of homology classes
\(i\) and \(j\), without considering self-interactions of classes.
The normalization factor in the denominator is chosen to be a linear
function of the number of edges.  The linear relation is suggested by
the experimental observation that the number of edges in these TF-TF
networks is comparable to the number of nodes.  We introduced the
square root normalization factor in order to compensate for the bias
generated by the different size of classes \(i\) and
\(j\)~\footnotemark[1].

\footnotetext[1]{The important fact here is that the normalization
  factor should compare to the number of nodes, as does the total
  number of edges in the network, alternative choices such as
  $C_i+C_j$ or $ sqrt(C_iC_j)$ lead to the same results.  }

\begin{table}
\centering
 \begin{tabular}{|c|c|c|c|}
 \hline
 \(R\) \TOPFIX\BOTTOMFIX & Empirical &  Randomization  & Z-score\\
 \hline
\emph{E.~coli} (RDB 5.5) & \(9.87\)   & \(7.71 \pm 3.65 \)    & \( +0.59 \) \\
\emph{B.~subtilis}    &  \(3.32\)     & \(3.46 \pm 2.64 \)    & \(\approx 0\) \\
\emph{S.~cerevisiae} (Balaji) & \(15.63\) & \(10.72 \pm 2.05 \) & \( +2.39 \)
\\ 
\hline
\end{tabular}
\caption{Rewiring parameter \(R\) for the TF-TF networks. For the
  bacterial datasets this observable falls in the typical case of
  a randomized instance. In the case of \emph{S.~cerevisiae}
  there is a significant trend which indicates absence of rewiring
  (positive Z score), despite of the  larger degree of feedback.
  This indicates that the feedback circuits
  tend to lie within homology classes.
  The parameter \(R\) quantifies the strength of asymmetric 
  regulations between homology classes and is defined in the text.}
\label{table:rew}
\end{table}

Note that the definition the collapsed network absorbs the presence of
crosstalks or feedback likely inherited by AR duplication, and this is
reflected by \(R\). 
To understand this, let us assume that the probability of a rewiring
event is smaller than the probability of keeping inherited
interactions. Then, homology classes, generated by duplication of
ancestral genes, would tend to maintain their inherited regulations,
thus measuring an high value of \(R\). In other words, \(R\) measures
the extent to which feedback is retained within homology classes.
This likely comes from AR duplication, or duplication of existing
higher-order feedbacks between homologs.

To sum up, the larger is \(R\) (compared to randomizations), the less
likely rewiring has shaped the network, and the more the hierarchy
between the homology classes is maintained.  Evaluation of empirical
data (Table~\ref{table:rew}) shows that in the bacterial data sets,
the estimated rewiring is consistent with, or slightly lower than,
typical randomized values.  Perhaps unexpectedly, we find an even
stronger signal for the yeast TF-TF network, where, as we have shown,
the feedback is more common. Most of this feedback involves homolog
TFs, and the network is likely to have been shaped by a small amount
of rewiring, or by rearrangements that keep into account the homology
relations of transcription factors.
Note that, empirically, one alternative possibility is that a
duplication of ARs with inherited crosstalk is followed by a
subsequent split the homology class, so that the two homologs cease to
be classified as such. This phenomenon would lead to a false positive
for rewiring in our interpretation. However, this problem is less
relevant in presence of a negative signal, such as the one we find.
Another way to put the same observation, is that most of the feedback
observed in yeast was shaped by duplication with retained crosstalks,
rather than by rewiring.

\paragraph*{Duplication of Homodimeric and Heterodimeric TFs.}

An issue where we found notable evolutionary differences between the
evolutionary transcriptional architecture of \emph{E.~coli} and
\emph{S.~cerevisiae} is the duplication of homo- and heterodimeric
TFs\footnotemark[2].

\footnotetext[2]{This analysis was not possible for \emph{B.~subtilis} due
to the lack of large-scale interaction data.}

It is known from sparse observations that yeast tends to use more
heterodimeric TF pairs than prokaryotes~\cite{IYM+05, LPC+06, PLK+07}.
A systematic analysis using large protein interaction datasets
confirms this trend.  In \emph{E.~coli} (out of 150 TFs) we find 21
homodimers, and only 5 heterodimer pairs, all formed with the
histone-like protein HU, which has a rather special status. Though no
systematic data is available, the common opinion is that it likely
that these figures strongly underestimate the number of homodimers. A
computational evaluation of the TFs available in the 3DComplex
database~\cite{LPC+06} is not able to enrich the sample. On the
other hand, it shows
that 37/42 TF entries in PDB are scored as homodimers, corroborating
the common belief.  Conversely, in yeast, where more systematic data
is available, we find (out of 157 TFs) 45 homodimers, and 91
heterodimer pairs.

\begin{table}
\centering
 \begin{tabular}{|c|c|c|c|c|}
 \hline
 & Co-occurrence\TOPFIX\BOTTOMFIX & Empirical & Randomization  & Z-score\\
 \hline
\multirow{2}{*}{A} & \emph{E.~coli} & \(0\) & \( 0.46\pm1.22 \) & \(+0.37\) \\
& \emph{S.~cerevisiae} & \(84\) & \( 32.38\pm18.95 \)  & \( +2.72\) \\
 \hline
\multirow{2}{*}{B} & \emph{E.~coli} & \(0\) & \( 0.213\pm0.45 \) & \(+0.47\) \\
& \emph{S.~cerevisiae} & \(15\) & \( 4.86\pm2.29 \) &  \( +4.43 \) \\
\hline
 \end{tabular}
 \caption{A. Homodimer Heterodimer TFs co-occurrence in
   classes measured by the product of the number of homodimers and
   heterodimers in each class, summed over classes.
   B. Homologous Heterodimer TFs in the same class. }
\label{table:heterod}
\end{table}

Inspection of the homology classes (Table \ref{table:heterod} and
Supplementary Figure) shows
that in \emph{S.~cerevisiae}, heterodimeric TFs tend to cluster
in homology classes, and also to co-occur with homodimers. We
interpret this as a strong indication that heterodimers stem from
duplications of other dimeric TFs.  This is measured by evaluating
the overrepresentation of number of homodimers and heterodimers in
homology classes, and the product of homodimers and heterodimers
in the same class.
The same process seems to be forbidden in \emph{E.~coli}, possibly because of
the same selective pressure erasing crosstalks from duplicate ARs. Due
to insufficient data, we could not show that homodimeric TFs in his
bacterium are likely to form classes of homodimers.

\paragraph*{Longest Path and Shortest Path Hierarchy.}

Finally, we proceed to evaluate the \emph{long} and \emph{short}
path hierarchy conservation in evolution. The model predicts that
they will be conserved, in absence of rewiring and multiple  AR
duplications. In particular, we expect to find many members of
the same homology class falling in the same computational
layer, defined by longest paths  to a root or shortest paths.
Our observations are reported in Table \ref{table:hierarchy}.
In \emph{E.~Coli} both data sets indicate a strong tendency to have duplicates
in the same layer, with both definitions of hierarchy.
In \emph{B.~subtilis} we find a similar trend only for the long-path
hierarchy (the signal is much weaker, possibly due to the
small size of the sample).  Finally, in yeast we also find evidence for
hierarchy conservation all data sets.  In the larger dataset, the
long-path hierarchy was evaluated for the hierarchical component only.

\begin{table}
\centering
 \begin{tabular}{|c|c|c|c|c|}
 \hline
Path\TOPFIX\BOTTOMFIX & Data Set & Empirical & Randomization  & Z-score\\
 \hline
\multirow{5}{*}{A - Long} & \emph{E.~coli}  (Shen-Orr) & \( 844 \)
 & \( 687.37 \pm 48.18 \) & \( +3.25 \)   \\

 & \emph{E.~coli}  (RDB 5.5)    & \( 1171\)
 & \( 420 \pm 34.7 \) & \(+21.6\)   \\

 & \emph{B.~subtilis}  & \(377\)
 & \( 353.43 \pm 22.81 \)  & \(+1.03\) \\

 & \emph{S.~cerevisiae} (Guelzim)&  \(290\)
 & \( 260 \pm 24.7 \)  & \(+1.21\) \\

 & \emph{S.~cerevisiae} (Balaji / TF)& \(299\)
 & \(262.88 \pm 20.80\) &  \(1.74\) \\
 \hline
%
%
\multirow{5}{*}{B - Short} & \emph{E.~coli} (Shen-Orr)& \( 1422 \)
 & \( 1367.3  \pm  15.79 \) & \(+3.5\) \\

 & \emph{E.~coli} (RDB 5.5) & $1193$
 & \( 453.69 \pm 40.88 \)  & \( +18 \) \\

 & \emph{B.~subtilis}       & \( 676 \)
 & \( 654.04 \pm 27.15 \)  & \( +0.19 \) \\

 & \emph{S.~cerevisiae} (Guelzim)&  \(365\)
 & \( 296.35 \pm  25.76\) & \(+2.67\) \\

 & \emph{S.~cerevisiae} (Balaji / TF)& \( 572\)
 & \( 543.97 \pm 18.25 \) & \( +1.54\) \\
\hline
 \end{tabular}
 \caption{Hierarchy and gene duplications. Preservation of the layer
   structure, measured  by the  number of homolog pairs occupying the same
   layer. (A) Longest path. (B) Shortest path (data for the Balaji data-set
   refer to the TF-TF interaction subnetwork).}
\label{table:hierarchy}
\end{table}

\section{Discussion and Conclusions}

Our results highlight both common and distinct trends in the two
bacteria and the yeast, both in the topology and in its evolution.  We
find that the underrepresentation of feedback is common in the three
transcription networks. Indeed, the feedback core is always smaller
than what expected from the null model, in both the two bacteria and
yeast. On the other hand, there is a difference in the size of the
feedback core of the network, which is 6 times larger in yeast,
involving a major fraction of the transcription factors. As noted by
Jeong and Berman~\cite{JB08}, this feedback essentially condensed in
one single connected component, and is enriched with TF nodes having
the endogenous functions of cell cycle and sporulation, and the
exogenous functions of diauxic shift and DNA repair. However, it must
be noted that this feedback component, though large, is much smaller
than expected by a degree-sequence-conserving null network model, so
that the trend of underrepresentation of feedback has to be regarded
as a common feature of the networks, and is particularly strong in
yeast.
On the other hand, there is a remarkable difference in the
hierarchical organization between the two evolutionarily distant
bacteria and yeast, which emerges more prominently in the length of
shortest paths in the network.  In the bacterial networks, the nodes
are organized in a small number of hierarchical layers pointing to a
minimization of both longest and shortest paths between TFs and their
targets. Since each computational layer has a cost in
time~\cite{REA02}, this observation would be consistent with a
possible constraint on the minimization of the time-scales for the
production of structural target genes from upstream TFs which sense a
particular internal or external signal.
In yeast, shortest paths are much longer, both in absolute terms and
compared to random networks. This means that the number of
computations made after an upstream TF is activated by a signal, so
that the hypothetic time constraints should be softer. This feature
could be ascribed to the documented more complex post-transcriptional
regulation and the efficient specific degradation machinery.
We also would like to observe that, while the distributions of small
network motifs seem to be common among known transcription
networks~\cite{MIK+04}, the nonlocal observables considered here,
which evaluate the feedback and hierarchichal organization of
transcription programs, point to consistent differences in network
architecture that accompany the transition from prokaryotes to
eukaryotes.

From the evolutionary analysis, we find in all cases significant
indications that the existing feedback stems from lower-order feedback
by gene duplication and inheritance of interactions. In particular, AR
duplication is significant for all data sets.  However, circuits with
crosstalks or feedback stemming from AR duplication are found in yeast
and to a certain extent in \emph{B.~subtilis}, but less in
\emph{E.~Coli}.
On the contrary, we observe that crosstalk conservation in
\emph{E.~coli} is not frequent.  Direct measurement of ARs in
\emph{E.~coli} show a consistent signal that the population of ARs in
homology classes is more dense and more variable than in randomized
instances~\cite{CLJ+07}. This is a direct evidence of duplication and
conservation of ARs during evolution.  In \emph{B.~subtilis} and
yeast, the population of ARs falls closer to the null model. The pure
analysis of AR population in homology classes is not sufficient to
assess that ARs evolve from duplication and inheritance of self-edges.

The other face of the medal is that outside of the small feedback
cores, a hierarchical organization where computational layers are
built by gene duplication is visible.  more in detail, in
\emph{E.~Coli} and \emph{S.~cerevisiae} duplicates tend to populate
the same layer, indicating conservation of hierarchy. The same seems
to be true for \emph{B.~subtilis}, but the signal is weaker.
A possible interpretation is again that, in bacteria, minimization
constraints on the time-scales for the production of target genes, if
present, would translate into selective pressure for the reduction of
the number of computational layers. On the other hand, such pressure
could be released from yeast due to more efficient
post-transcriptional control of gene expression.

Finally, and along the same lines, we find a strong distinct trend in
\emph{E.~coli} and yeast, concerning the use and evolution of
homodimers \emph{versus} heterodimer transcription factors.  In
\emph{S.~cerevisiae} we found strong indication that heterodimers TFs
stem from duplications of ancestral dimeric TFs.  The same process
seems to be forbidden in \emph{E.~coli}, possibly because of the same
selective pressure erasing crosstalks from duplicate ARs.
One can speculate that this ability to make use of heterodimeric
binding boosts the combinatorial capacity of a promoter signal
integration function~\cite{BBG+05}.

In conclusion, we presented a comparative analysis of large-scale
topological and evolutionary properties of transcription networks in
three species, focusing on the global aspects of feedback and
hierarchy in regulatory pathways.
This analysis confirms that gene duplication is an important drive for
the shaping of transcription networks, which follows distinct
directions between bacteria, where we hypothesize time constraints to
impose the observed simple hierarchical structure, and yeast, where
more intricate pathways arise.  Overall, it appears that yeast is
able to sustain a higher complexity in its topological structure,
including more feedback and longer pathways, and to explore more
freely the possible regulatory interactions stemming from gene
duplication, such as feedback stemming from self-interactions and
dimer transcription factors.  Our current work explores the hypothesis
that part of this higher complexity stems from the known whole-genome
duplication event~\cite{WPF+07,WT07}.


\section*{METHODS}
\label{sec:methods}

\paragraph*{Graph Growth Model.}
A simple model of network evolution through duplication-divergence was
considered.  At each time step all the nodes of the graph are
duplicated, while the number of edges rises fourfold.  This happens
for the following reason: for each edge connecting two original (old)
nodes (\emph{old-old} edge), duplication of interaction gives rise to
edges between the two old nodes and the two duplicate nodes. The
original \emph{old-old} edge therefore generates the four
\emph{old-old}, \emph{old-new}, \emph{new-old}, \emph{new-new} edges.
Duplication of the graph is followed by erasing of edges with
prescribed probabilities~\cite{tesikirill,CLJ+07}.  One can formulate
the model with partial or global duplications, and including or not
the duplication and removal of self-edges (in this case, we considered
the probability of retaining a self-edge equal to that of any other
edge).  The behavior of this model was compared, through a set of
observables, with the observed trends of the experimental data.

\paragraph*{Data Sets.}
We considered the following data sets for the transcription networks.
For \emph{E.~coli}, the Shen-orr data-set and the larger and more
recent RegulonDB5.5~\cite{SMM+02,SGP+06,SSG+06}.  For
\emph{B.~subtilis}, DBTBS~\cite{MNO+04}.  For \emph{S.~cerevisiae},
the Guelzim~\cite{GBB+02} data-set and the more recent
Balaji~\cite{BBI+06} data set. The Balaji data set was modified to
include auto regulating interactions taken from the literature. In the
case of yeast, probably due to the much larger size of the more recent
data, there is no compatibility between the two sets of data.
Domain architecture data are taken from the SUPERFAMILY
database~\cite{GKH+01, WMV+07}, versions 1.61 and 1.69, as in the data
sets in~\cite{BLA+04}.
Homodimers and Heterodimers for\emph{S.~cerevisiae} and \emph{E.~coli}
respectively, were obtained from the SGD database~\cite{CWB+04} and
from ref.~\cite{BPL+05}.

\paragraph*{Evaluation of Feedback and Hierarchy.}
We used the leaf-removal algorithm~\cite{CLJ+07} on the data-sets
(including ARs) and their randomized counterparts. This algorithm
prunes the input and output tree-like components of a directed network
leaving with a ``feedback core'' of nodes, where each node is involved
in at least one feedback loop.  Each iteration of this pruning
algorithm defines a hierarchical layer. The main observables we
considered were the size of the core and the number of iterations to
reach the core.
For the evaluation of shortest paths, we used the Dijkstra
algorithm~\cite{Dij59}, considering the distribution of shortest-path
lengths and the longest paths.  The results for the empirical networks
were compared to randomizations obtained using a standard Markov Chain
Monte Carlo (MCMC) algorithm~\cite{RJB96} or an Importance Sampling
Monte Carlo algorithm~\cite{FBJ+07} that preserve the degree sequence
(marginals of the adjacency matrix).  We considered random
counterparts of the networks where the only constraints come from the
in- and out- degree sequences. In particular, we chose not to conserve
the number of self-regulators. Our previous work~\cite{CLJ+07,FBJ+07}
shows that in general this null model yields qualitatively different
results. In particular, in the case of the Shen-Orr data set the
number of layers falls in the average of the random ensemble that
conserves self-regulations.  For the scopes of this work, we did not
consider this kind of randomization, and we hypothesized that, as in
the model for the graph growth by duplication, auto-regulatory edges
have the same status of other edges.

\paragraph*{Evaluation of Duplications.}
We used domain architectures from the SUPERFAMILY database to build
protein architecture databases, one for each specie. We then
constructed classes of homologous genes using similarity criteria
between these architectures, as was done in~\cite{TB04}.  Two genes
are considered homologs if they share the same domains in the same
order, neglecting domain repeats. For this analysis, proteins coded by
the same operon were considered as separate entities.  Since the
definition of homology is rather arbitrary it is rather natural to
test different definition and observe the stability of results. All
these results were filtered for consistency using stricter or looser
homology criteria~\cite{sellerio002}.  In the case of the
\emph{S.~cerevisiae}, and only with respect to the network degree
distributions (as shown in Fig.~\ref{fig:degree}), we also considered
the notion of homology descending from the gene pairs defined by
blocks of doubly conserved syntheny described in~\cite{KBL04}.

\paragraph*{Duplication of Auto Regulators.}
We assessed the evolution of the Auto Regulated transcription factors
by studying the distribution of the subgraphs shown in
Fig.~\ref{fig:arsub} within homology classes.  Quantitative measure
was given by the evaluation of observable \(a\), as described in the
main text, which was computed on both the experimental data set and on
the randomized instances of the null model described later.

\paragraph*{Interaction and Rewiring between TFs.}
We studied the distribution of regulating interaction within and
across homology classes.  A number of observables were implemented to
describe the relationship standing between the architecture data set
and the transcription network data set.  To quantify the effect of
rewiring we introduced the observable \(R\) as described in main text.
Again, this observable was evaluated upon the experimental data set
and upon the null model.

\paragraph*{Evaluation of Coevolution of Dimer TFs.}
We studied the evolutionary properties of homodimers and heterodimers
with respect to our duplication and divergence model.  We assessed the
co-existence of homo- and hetero- dimers within the homology classes
by evaluating the sum (over all the homology classes) of the number of
all dimeric pairs found in each homology class.  The same analysis
were performed upon the randomized instances of the null model, and
the result is shown in Fig.~\ref{fig:hhsumproduct}.

\paragraph*{Null Model for Duplication.}
Most of the measures performed upon the experimental data sets were
compared with a null model of homology.  Keeping fixed the homology
classes, we randomly shuffled the architecture associations to the
gene names.  This randomizes all the interaction existing between
homology properties (which are responsible of class generation) and
all the other data sets (adjacency list of transcription network,
homodimers, heterodimer pairs).  The advantage of this null model lies
in the fact that no experimental database is actually randomized, only
their interactions. This is important as this randomization does not
destroy the homology information nor the global (large scale)
properties of the network.  Finally, gene name shuffling was done
separately for TFs and TGs, as was done in~\cite{TB04}, due to their
inherently different DNA-binding properties (which depend on their
domains).  The data shown in this paper correspond to \(10^{5}\)
randomized instances of the null model, allowing us to estimate
P-values larger than \(1 \times 10^{-5}\).

\section*{Acknowledgments.}
We would like to thank Kirill Evlampiev for useful discussions and
work on the asymptotics of the theoretical model, Diana Fusco for help
with the subgraph analysis.






\newpage

\section*{SUPPLEMENTARY MATERIAL}

\renewcommand{\thesection}{S\arabic{section}}
\setcounter{figure}{0} 
\setcounter{table}{0} 
\setcounter{section}{0} 
\renewcommand{\figurename}{Supplementary Figure}
\renewcommand{\thefigure}{\arabic{figure}}
\renewcommand{\tablename}{Supplementary Table}
\renewcommand{\thetable}{\arabic{table}}

\newpage

\begin{figure}
  \centering
  \includegraphics[width=0.5\textwidth]{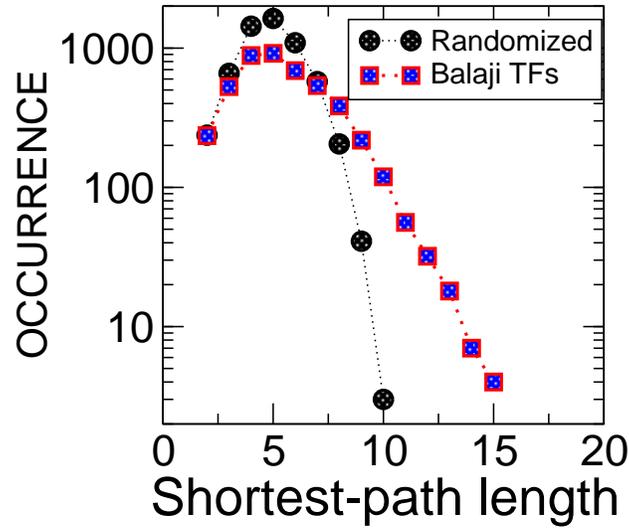}
  \caption{Histogram of shortest-path lengths in the yeast TF network
    (Balaji data set), compared to a typical randomized instance. The
    graph plots the number of shortest paths of a given length.  The
    two distributions have similar qualitative features, in particular
    showing exponential decay and the same location of the
    maximum. However, they differ significantly in their tails, where
    the empirical graph shows a higher abundance of longer
    shortest-paths.  }
  \label{fig:yeast_shpdist}
\end{figure}

\newpage

\begin{table}
\centering
 \begin{tabular}{|c|c|c|c|}
 \hline

 \emph{E.~coli} (RDB 5.5)   &     &     &    \\

 Subgraph type \TOPFIX\BOTTOMFIX & \# Empirical &  Randomization & P-value\\

 \hline
\textbf{(i)}   & \(1\)  & \( 7.877  \pm  3.605 \)   & \(0.0017\) \\
\textbf{(ii)}   &  \(3\)  & \( 0.412  \pm  0.640 \)   & \(0.0087\) \\
\textbf{(iii)}   &  \(6\)  & \( 4.089  \pm  2.639 \)   & \(0.2497\) \\
\textbf{(iv)}   &  \(1\)  & \( 0.1072  \pm  0.319 \)   & \(0.1042\) \\
\textbf{(v)}    &  \(0\)  & \( 0  \pm 0 \)   & \textit{n.a.} \\
\textbf{(vi)}   &  \(179\)  & \( 81.55  \pm  15.24 \)   & \(<10^{-4}\) \\
 \hline

\end{tabular}

\vspace{1cm}

 \begin{tabular}{|c|c|c|c|}
 \hline

\emph{B.~subtilis} (DBTBS)  &     &     &    \\

  Subgraph type \TOPFIX\BOTTOMFIX & \# Empirical  &  Randomization & P-value\\

\hline
\textbf{(i)}   & \( 1 \)  & \( 1.2472  \pm 1.0994  \)   & \(  0.2824  \) \\
\textbf{(ii)}   &  \( 2 \)  & \(  1.7739 \pm 1.8284  \)   & \(  0.4327 \) \\
\textbf{(iii)}   &  \( 1 \)  & \( 0.2370  \pm  0.4872 \)   & \(  0.2101 \) \\
\textbf{(iv)}   &  \( 1 \)  & \( 0.0755 \pm 0.2642 \)   & 0.0755 \\
\textbf{(v)}    &  \( 0 \)  & \( 0.0758  \pm 0.2646 \)   & \( 0.0758  \)  \\
\textbf{(vi)}   &  \( 55 \)  & \(  29.237 \pm  10.672 \)   & \(  0.0242  \) \\
\hline

\end{tabular}

\vspace{1cm}

\begin{tabular}{|c|c|c|c|}
  \hline

\emph{S.~cerevisiae} (Balaji)   &     &     &    \\
  
  Subgraph type \TOPFIX\BOTTOMFIX & \# Empirical  &  Randomization & P-value\\
  
\hline
\textbf{(i)}   & \(7\)  & \( 8.756  \pm  3.991 \)   & \(0.3074\) \\
\textbf{(ii)}   &  \(6\)  & \( 2.373  \pm  1.435 \)   & \(0.0243\) \\
\textbf{(iii)}   &  \(0\)  & \( 0.209  \pm  0.414 \)   & \(<10^{-4}\) \\
\textbf{(iv)}   &  \(0\)  & \( 0 \pm 0 \)   & \textit{n.a.} \\
\textbf{(v)}    &  \(1\)  & \( 0.199  \pm 0.399 \)   & \( 0.1995 \)  \\
\textbf{(vi)}   &  \(16\)  & \( 17.49  \pm  6.51 \)   & \( 0.4308 \) \\
\hline

\end{tabular}
\caption{Subgraphs compatible with AR duplications, containing
  crosstalks. The tables report  the occurrence of
  different network subgraphs which  can stem from duplications of
  nodes and edges and subsequent partial loss of 
  regulatory interactions for the three transcription networks,
  compared to randomizations of the homology classes. These subgraphs 
  are displayed in Fig.~\ref{fig:arsub}.} 
\label{table:subgraphs_AR}
\end{table}

\newpage

\begin{figure}[htbp]
  \centering
  \includegraphics[width=0.7\textwidth]{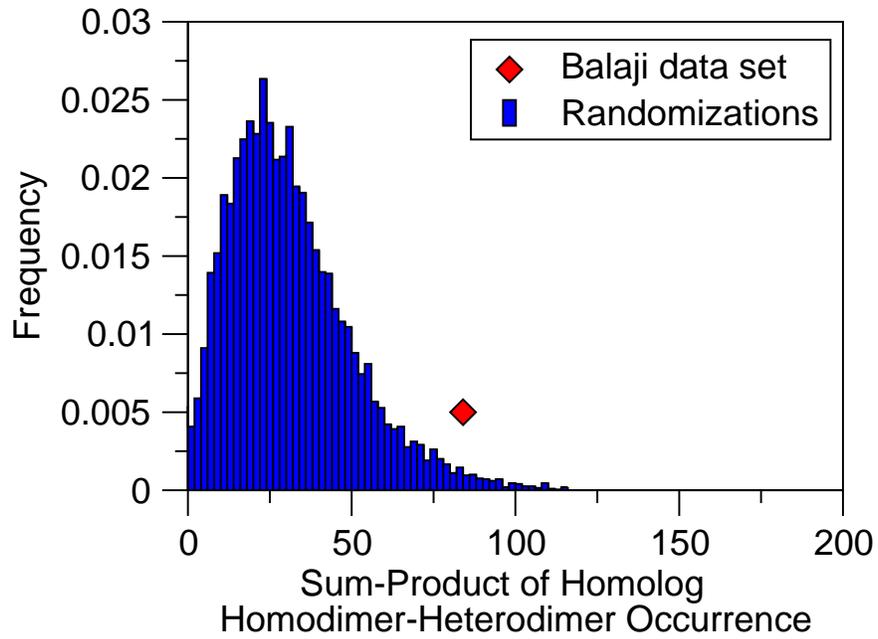}
  \caption{Heterodimer / homodimer TF duplication in yeast.  The x
    axis reports the observable quantifying the co-occurrence of
    homodimers and heterodimers in the same homology class (the sum
    over classes of the product of number of heterdimers and
    homodimers within each class), for the empirical network of yeast
    (Balaji data-set), compared to randomizations (histogram).  The
    histogram corresponds to the data shown in
    Table~\ref{table:heterod}.}
  \label{fig:hhsumproduct}
\end{figure}

\end{document}